\documentclass{aa}

\usepackage[varg]{txfonts}
\usepackage{epsfig,graphicx,natbib,url,twoopt}
\usepackage[varg]{txfonts}
\usepackage{hyperref}          %% for pdflatex
\usepackage{float}
\hypersetup{
 colorlinks=true,  %RR colors links instead of ugly boxes
 urlcolor=blue,    %RR color external hyperlinks
 linkcolor=red,     %RR color internal links
 citecolor=blue %% color of citations, default green is almost invisible
}
\def\kms{\hbox{km$\;$s$^{-1}$}}

\def\Halpha{\mbox{H\hspace{0.1ex}$\alpha$}}

\def\FeI{\ion{Fe}{i}}
\def\CaIR{\ion{Ca}{ii}~8542}
\def\CaK{\ion{Ca}{ii}~K}
\def\CaH{\ion{Ca}{ii}~H}

\begin{document}

\title{High-resolution observations of the solar photosphere, chromosphere, and transition region}
\subtitle{A database of coordinated IRIS and SST observations}

\author{L.H.M.~Rouppe van der Voort\inst{\ref{inst:ITA},\ref{inst:RoCS}}
\and
B.~De~Pontieu\inst{\ref{inst:LMSAL},\ref{inst:ITA},\ref{inst:RoCS}} % Bart - ok+
\and
M.~Carlsson\inst{\ref{inst:ITA},\ref{inst:RoCS}} % Mats - ok+
\and
J.~de~la~Cruz~Rodr{\'i}guez\inst{\ref{inst:Stockholm}} % Jaime - ok+
\and
S.~Bose\inst{\ref{inst:ITA},\ref{inst:RoCS}} % Souvik - ok+
\and
G.~Chintzoglou\inst{\ref{inst:LMSAL},\ref{inst:UCAR}} % Georgios - ok+
\and
A.~Drews\inst{\ref{inst:ITA},\ref{inst:RoCS}} % Ainar - ok+
\and
C.~Froment\inst{\ref{inst:ITA},\ref{inst:RoCS},\ref{inst:orleans}} % Clara - ok+
\and
M.~Go\v{s}i{\'c}\inst{\ref{inst:LMSAL},\ref{inst:BAERI}} % Milan Gošić - ok+
\and
D.R.~Graham\inst{\ref{inst:LMSAL},\ref{inst:BAERI}} % David - ok+
\and
V.H.~Hansteen\inst{\ref{inst:ITA},\ref{inst:RoCS},\ref{inst:LMSAL},\ref{inst:BAERI}} % Viggo - ok+
\and
V.M.J.~Henriques\inst{\ref{inst:ITA},\ref{inst:RoCS}} % Vasco - ok+
\and
S.~Jafarzadeh\inst{\ref{inst:ITA},\ref{inst:RoCS}} % Shahin - ok+
\and
J.~Joshi\inst{\ref{inst:ITA},\ref{inst:RoCS}} % Jayant - ok+
\and
L.~Kleint\inst{\ref{inst:UASANS},\ref{inst:KIS}} % Lucia - ok+
\and
P.~Kohutova\inst{\ref{inst:ITA},\ref{inst:RoCS}} % Petra - ok+
\and
T.~Leifsen\inst{\ref{inst:ITA},\ref{inst:RoCS}} % Torben - ok+
\and
J.~Mart{\'i}nez-Sykora\inst{\ref{inst:LMSAL},\ref{inst:BAERI},\ref{inst:ITA},\ref{inst:RoCS}} % Juan - ok+
\and
D.~N{\'o}brega-Siverio\inst{\ref{inst:ITA},\ref{inst:RoCS}} % Daniel - ok+
\and
A.~Ortiz\inst{\ref{inst:ITA},\ref{inst:RoCS}} % Ada - ok+
\and
T.M.D.~Pereira\inst{\ref{inst:ITA},\ref{inst:RoCS}} % Tiago - ok+
\and
A.~Popovas\inst{\ref{inst:ITA},\ref{inst:RoCS}} % Andrius - ok+
\and
C.~Quintero~Noda\inst{\ref{inst:ITA},\ref{inst:RoCS}} % Carlos - ok+
\and
A.~Sainz~Dalda\inst{\ref{inst:LMSAL},\ref{inst:BAERI},\ref{inst:stanford}} % Alberto - ok+
\and
G.B.~Scharmer\inst{\ref{inst:Stockholm}} % Goran -ok+
\and
D.~Schmit\inst{\ref{inst:Catholic}} % Don - ok+
\and
E.~Scullion\inst{\ref{inst:Newcastle},\ref{inst:ITA}} % Eamon - ok+
\and
H.~Skogsrud\inst{\ref{inst:ITA}} % H{\aa}kon - ok+
\and
M.~Szydlarski\inst{\ref{inst:ITA},\ref{inst:RoCS}} % Mikolaj - ok+
\and
R.~Timmons\inst{\ref{inst:LMSAL}} % Ryan - ok
\and
G.J.M.~Vissers\inst{\ref{inst:Stockholm},\ref{inst:ITA}} % Gregal - ok+
\and
M.M.~Woods\inst{\ref{inst:LMSAL},\ref{inst:BAERI}} % Magnus - ok+
\and
P.~Zacharias\inst{\ref{inst:ITA}}% Pia - ok+
}

\authorrunning{Rouppe van der Voort et al.}

\institute{Institute of Theoretical Astrophysics,
  University of Oslo, %
  P.O. Box 1029 Blindern, N-0315 Oslo, Norway
  \label{inst:ITA}
\and
 Rosseland Centre for Solar Physics,
  University of Oslo, %	
  P.O. Box 1029 Blindern, N-0315 Oslo, Norway
  \label{inst:RoCS}
\and 
  Lockheed Martin Solar \& Astrophysics Laboratory, 3251 Hanover St., Palo Alto, CA 94304, USA
  \label{inst:LMSAL}
\and
  Institute for Solar Physics, Dept. of Astronomy, Stockholm University, AlbaNova University Centre, 10691, Stockholm, Sweden
  \label{inst:Stockholm}
\and
  University Corporation for Atmospheric Research, Boulder, CO 80307-3000, USA
  \label{inst:UCAR}
\and 
  LPC2E, CNRS and University of Orl{\'e}ans, 3A avenue de la Recherche Scientifique, Orl{\'e}ans, France 
  \label{inst:orleans}
\and 
  Bay Area Environmental Research Institute, NASA Research Park, Moffett Field, CA 94035, USA
  \label{inst:BAERI}
\and
  University of Applied Sciences and Arts Northwestern Switzerland, Bahnhofstrasse 6, 5210 Windisch, Switzerland
  \label{inst:UASANS}
\and
  Leibniz-Institut f{\"u}r Sonnenphysik (KIS), Sch{\"o}neckstrasse 6, D-79104 Freiburg, Germany
  \label{inst:KIS}
\and
  Stanford University, HEPL, 466 Via Ortega, Stanford, CA 94305-4085
  \label{inst:stanford}
\and
  Catholic University of America, Washington, DC 20064, USA; NASA Goddard Space Flight Center, Greenbelt, MD 20771, USA\label{inst:Catholic}
\and
  Department of Mathematics, Physics and Electrical Engineering, Northumbria University, Newcastle upon Tyne NE1 8ST, UK
  \label{inst:Newcastle}
  }

\date{\today\ - accepted for publication in A\&A July 21, 2020}

% \abstract{}{}{}{}{} 
% 5 {} token are mandatory
% "Aims", "Methods", and "Results", are mandatory
\abstract
{NASA's Interface Region Imaging Spectrograph (IRIS) provides high-resolution observations of the solar atmosphere through ultraviolet spectroscopy and imaging. 
Since the launch of IRIS in June 2013, we have conducted systematic observation campaigns in coordination with the Swedish 1-m Solar Telescope (SST) on La Palma.
The SST provides complementary high-resolution observations of the photosphere and chromosphere.
The SST observations include spectropolarimetric imaging in photospheric \ion{Fe}{i} lines and spectrally resolved imaging in the chromospheric \ion{Ca}{ii}~8542~\AA, \Halpha, and \ion{Ca}{ii}~K lines.
We present a database of co-aligned IRIS and SST datasets that is open for analysis to the scientific community. 
The database covers a variety of targets including active regions, sunspots, plages, the quiet Sun, and coronal holes.
}
%{% Context
%Database
%}
%{% Aims
%We aim to 
%}
%{% Methods
%We analyze high spatial (0\farcs1) resolution  observations obtained with the Swedish 1-m Solar Telescope.
%}
%{% Results
%}
%{% Conclusions
%}
% abstract: limit to 300 words, and be self-contained

% Sun: faculae, plages -- Sun: flares -- Sun: UV radiation -- sunspots

\maketitle

%%%%%%%%%%%%%%%%%%%%%% Introduction %%%%%%%%%%%%%%%%%%%%%%%
\section{Introduction}
\label{sec:introduction}

The solar atmosphere is a very dynamic region, where fundamental physical processes take place on small spatial scales and short dynamical time scales, often leading to rapid changes in the thermodynamic state of the plasma.
Resolving these processes in observations requires high resolution in the combined spatial, temporal, and spectral domains.
Furthermore, the combination of multiple spectral diagnostics, preferably with sensitivity to line formation conditions that cover a large range in temperatures, densities, and magnetic field topologies, are of fundamental importance for advancing our understanding of the solar atmosphere.
The simultaneous acquisition of vastly different spectral diagnostics is possible through coordinated observations between space-borne and ground-based observing facilities.
Telescopes in space provide unique access to the short wavelength regime with seeing-free diagnostics of the chromosphere, transition region and corona. 
Ground-based telescopes allow for high resolution in photospheric and chromospheric diagnostics, as well as high-sensitivity polarimetric measurements of the magnetic field with instrumentation that can be more complex than in space, and which is not limited by data transfer rates.
Coordinated observations, therefore, strongly enhance the potential to unravel connections in the
solar atmosphere that span from the photosphere, through the chromosphere and transition region to the corona. 

The Interface Region Imaging Spectrograph 
\citep[IRIS, ][]{2014SoPh..289.2733D}, % BDP IRIS
a NASA Small Explorer (SMEX) satellite, was launched on 2013-Jun-27. %June 27, 2013.
It combines high resolution in the spatial (0\farcs3--0\farcs4), temporal (down to 1~s), and spectral domains (velocity determination down to 1~\kms).
Spectral diagnostics include the \ion{Mg}{ii}~h \& k resonance lines (chromosphere), the \ion{C}{ii} lines at 1335~\AA\ (upper chromosphere and transition region), and the \ion{Si}{iv} lines at 1400~\AA\ (transition region). 
Furthermore, the (weaker) \ion{O}{iv} lines around 1400~\AA\ as well as the \ion{Fe}{xii}~1349~\AA\ and \ion{Fe}{xxi}~1354~\AA\ lines provide diagnostics on the corona and high-energy flares. 
Slit-jaw imaging in the \ion{Mg}{ii}~k core, \ion{Mg}{ii}~h wing, \ion{C}{ii}, and \ion{Si}{iv} lines provides valuable context information.

The IRIS satellite offers considerable flexibility in its observing configuration, and, for example, allows for a wide variety in area coverage (i.e., field-of-view (FOV) size), temporal cadence, and choice of spectral diagnostics. 
Target selection is organized through a system with relatively short communication lines and allows for effective coordination with ground-based telescopes and other observing facilities.
This has opened up possibilities to expand on IRIS's rich arsenal of spectral diagnostics, for example by adding photospheric and chromospheric spectropolarimetry and high-resolution imaging in various spectral lines at and around the area covered by the IRIS spectrograph slit. 

Shortly after IRIS was launched, scientists from the University of Oslo and from the Lockheed Martin Solar and Astrophysics Laboratory (LMSAL) started organizing coordinated observing campaigns with the Swedish 1-m Solar Telescope
\citep[SST, ][]{2003SPIE.4853..341S}  % Scharmer+ SST
on La Palma.
Every year, four campaigns -- typically two weeks each -- are conducted during the SST observing season (April -- October).
The SST is capable of providing high-quality time series of spectrally resolved photospheric and chromospheric diagnostics that under excellent seeing conditions reach the diffraction limit of $<$0\farcs1 over the full arcmin$^2$ FOV. 
Furthermore, the versatile CRISP instrument can provide spectropolarimetric data that enable the measurement of the magnetic field topology. In addition, the tunable filter system CHROMIS, installed in 2016, can simultaneously provide narrowband filtergrams at several wavelengths in the core of the \ion{Ca}{ii}~K line.

Data from the coordinated campaigns have been used to study a variety of topics, including: the disk counterparts of spicules 
\citep{2014Sci...346D.315D, % BDP+ Science twist
2015ApJ...799L...3R, % Rouppe+ IRIS RBEs 
2017A&A...597A.138R, % Rutten contrails
2017Sci...356.1269M, % Juan+ spicules Science
2019A&A...631L...5B}, % Souvik+ REs in Ca K and Mg k
chromospheric bright grains in the internetwork
\citep{2015ApJ...803...44M} % Juan+ IRIS QS grains
and active region plage
\citep{2016ApJ...817..124S}, % Skogsrud+AR bright grains in SJI1400
penumbral microjets in sunspots
\citep{2015ApJ...811L..33V, % Vissers+ PMJs
2020A&A...638A..63D}, % Ainar & Rouppe PMJ
the atmospheric stratification in plage
\citep{2015ApJ...809L..30C} % Mats Mg k plage
and sunspots
\citep{2019A&A...627A..46B}, % Souvik+ sunspot atmosphere models
the relation between Ellerman Bombs and ultraviolet (UV) bursts
\citep{2015ApJ...812...11V, % Vissers+ EBs + UVBursts
2017ApJ...839...22H, % Viggo Bombs & Flares
2017ApJ...851L...6R, % Rouppe+ plasmoids
2019A&A...627A.101V, % Gregal+ CRISP+CHROMIS+IRIS EB/UVB inversions
2020A&A...633A..58O}, % Ada+ EBs + UVBs
Ellerman bombs in the quiet Sun
\citep{2016A&A...592A.100R}, % Rouppe+ QSEBs
magnetic flux emergence from the photosphere to the transition region
\citep{2016ApJ...825...93O}, % Ada Fluxem III
surges \citep{2017ApJ...850..153N}, % Daniel+ surge
and the chromospheric counterparts of transition-region unresolved fine structure loops
\citep{2018A&A...611L...6P}. % Tiago+ UFS

In this paper, we describe the public release of co-aligned IRIS and SST data. 
%L: did not incorporate LE "the co-aligned IRIS and SST data"
%
At first, the public release is limited to data products that share the same plate scale as IRIS (0\farcs17 per pixel) for easier data analysis.
This pixel scale implies that the spatial resolution of the SST data is degraded.
The release of the corresponding full spatial resolution SST data is planned for future data releases.

%%%%%%%%%%%%%%%%%%%%%% Observations %%%%%%%%%%%%%%%%%%%%%%%
\section{Observations and data processing}
\label{sec:obs}

\subsection{IRIS}
\label{sec:IRIS}

The IRIS telescope design and instrumentation are described in 
\citet{2014SoPh..289.2733D}. % BDP+ IRIS
The IRIS satellite acquires spectra in three spectral regions: in the far UV from 1332 to 1358~\AA\ (FUV1), in the far UV from 1389 to 1407~\AA\ (FUV2), and in the near UV from 2783 to 2834~\AA\ (NUV). 
The FUV1 region is dominated by the \ion{C}{ii} lines at 1334 and 1335~\AA\ that are formed in the upper chromosphere
\citep{2015ApJ...811...81R, 2015ApJ...814...70R}, % Rathore+ C II IRIS lines
the FUV2 region is dominated by the \ion{Si}{iv} lines at 1394 and 1403~\AA\ that are formed in the transition region.
The NUV region is dominated by the chromospheric \ion{Mg}{ii}~h and k lines
\citep{2013ApJ...772...89L,2013ApJ...772...90L}, % Leenaarts+ Mg h&k IRIS
and further hosts the upper photospheric and lower chromospheric \ion{Mg}{ii} triplet lines 
\citep{2015ApJ...806...14P} % Pereira+ Mg triplet
and a large number of (upper) photospheric blends in the strong \ion{Mg}{ii} wings 
\citep{2013ApJ...778..143P}. % Pereira+ Mg wing blends

The 0\farcs33 wide spectrograph slit has a length of 175\arcsec\ and can be displaced with respect to the solar surface to build up a raster that samples an area up to 130\arcsec$\times$175\arcsec.
There are several choices of step sizes between consecutive slit positions: dense sampling with 0\farcs35 steps, sparse sampling with 1\arcsec\ steps, or coarse sampling with 2\arcsec\ steps. Alternatively, the spectrograph can record data in a sit-and-stare mode, where the slit does not move and stays at a fixed location (with or without tracking for solar rotation). 

The IRIS satellite can take slit-jaw images (SJIs) with different filters to provide context around the spectrograph slit. %
The four science SJI channels are: SJI~2796, centered on \ion{Mg}{ii}~k (4~\AA\ bandpass); SJI~2832, centered at 2830~\AA\ in the \ion{Mg}{ii}~h wing (4~\AA\ bandpass); SJI~1330, centered at 1340~\AA\ and dominated by the \ion{C}{ii} lines (55~\AA\ bandpass); and SJI~1400, centered at 1390~\AA\ and dominated by the \ion{Si}{iv} lines (55~\AA\ bandpass).
Slit-jaw images from different channels are recorded sequentially and have the same exposure time as the spectrograms recorded with the spectrograph. 

Various choices can be made to reduce the data volume in order to fit within the daily limits of data transfer from the spacecraft to ground stations.
For each spectral line of interest, the wavelength range can be selected to limit the data transferred, or the spatial extent of the raster can be limited by transferring only data from a reduced part along the slit.
Other measures to reduce data transfer are compression, data binning (spatially and/or spectrally), and omitting one or several SJI channels (most frequently SJI 2832 is omitted, although this is often done to improve the cadence of the other SJI channels). 

Taken together, the various possible choices in raster step size, number of slit positions, slit length, SJI channel selection, exposure time, spatial and spectral binning, compression, and spectral line selection (line lists) constitute a considerable number of possible observing programs. 
These programs are identified by a unique number, the OBS number 
\citep[or OBSID; for more details, see ][]{2014SoPh..289.2733D}. % BDP+ IRIS 
The OBSID, together with the observing date and start time, constitute a unique identifier for each dataset (see the first three columns in Table~\ref{table:datasets}).

\subsection{SST}
\label{sec:SST}

The SST telescope design and its main optical elements are described in 
\citet{2003SPIE.4853..341S}. % Scharmer+ SST
A description of upgrades of optical components and instrumentation, as well as a thorough evaluation of optical performance is provided by
\citet{2019A&A...626A..55S}. % Scharmer+ Sky is limit
An adaptive optics system is fully integrated in the optical system 
\citep{2003SPIE.4853..370S} % Scharmer+ AO
and was upgraded with an 85-electrode deformable mirror operating at 2~kHz in 2013. 
A dichroic beam splitter divides the beam on the optical table into a red ($>500$~nm) and a blue beam. 
Both beams are equipped with tunable filter instruments: the CRISP imaging spectropolarimeter
\citep{2008ApJ...689L..69S} % Scharmer+ CRISP
on the red beam, and the CHROMIS imaging spectrometer on the blue beam. %
Both CRISP and CHROMIS are dual Fabry–P{\'e}rot filtergraph systems based on the design by 
\citet{2006A&A...447.1111S} % Scharmer FPI
and are capable of fast wavelength sampling of spectral lines. 
Before the installation of CHROMIS in September 2016, the blue beam was equipped with a number of interference filters, including a full width at half maximum (FWHM) of 10~\AA\ wide filter for photospheric imaging at 3954~\AA, and an FWHM=1~\AA\ wide filter centered on the \CaH\ line core at $\lambda=3968$~\AA\
\citep[see][]{2011A&A...533A..82L}. % Lofdahl+ Ca H filter on rotation table
The CRISP instrument has a pair of liquid crystals that together with a polarising beam splitter allow measurements of circular and linear polarisation in for example the photospheric \FeI~6173~\AA, \FeI~6301~\AA, and \FeI~6302~\AA\ lines, and the chromospheric \ion{Ca}{ii}~8542~\AA\ line. 

The CRISP instrument has a plate scale of 0\farcs058 per pixel and the SST diffraction limit ($\lambda/D$) is 0\farcs14 at the wavelength of \Halpha\ (with the telescope aperture diameter $D$=0.97~m). 
The transmission profile of CRISP has FWHM=60~m\AA\ at the wavelength of \Halpha. 
The CHROMIS instrument has a plate scale of 0\farcs038 per pixel and the SST diffraction limit is 0\farcs08 at the wavelength of \CaK.
The transmission profile of CHROMIS has FWHM$\approx$120~m\AA\ at the wavelength of \CaK. 
The FOV of CRISP and CHROMIS is approximately 1\arcmin$\times$1\arcmin. We note that sunlight collected by the SST is split by a dichroic beam splitter such that CRISP and CHROMIS can operate independently and in parallel, without reducing the efficiency of either instrument.

Image restoration by means of the multi-object multi-frame blind deconvolution 
\citep[MOMFBD, ][]{2002SPIE.4792..146L, % Lofdahl MFBD
2005SoPh..228..191V} % van Noort+ MOMFBD
method is applied to all data to enhance image quality over the full FOV.
The MOMFBD restoration is integrated in the CRISP and CHROMIS data processing pipelines
\citep{2015A&A...573A..40D, % De la Cruz Rodriguez CRISPRED
2018arXiv180403030L}. % Lofdahl+ CHROMISRED
These pipelines include the method described by 
\cite{2012A&A...548A.114H} % Vasco post-MOMFBD destretch
for consistency between sequentially recorded liquid crystal states and wavelengths, with destretching performed as in 
\cite{1994ApJ...430..413S}. % Shine destretch
The CRISP and CHROMIS instruments include auxiliary wideband (WB) systems which are essential as anchor channels in MOMFBD restoration. 
Furthermore, they provide photospheric reference channels that facilitate precise co-alignment between CRISP and CHROMIS data (or blue beam filter data before 2016), or co-alignment with data from IRIS and the Solar Dynamic Observatory
\citep[SDO,][]{2012SoPh..275...17L}. % Lemen+ AIA / SDO 

\subsection{SST observing programs}
\label{sec:SSTobs}

The SST observing programs vary from campaign to campaign, and often during campaigns as well, depending on the target and science goals. 
Common to all datasets in the database is the inclusion of at least one chromospheric line, \Halpha\ or \CaIR~\AA, and often both. 
In order to keep the temporal cadence below 20~s, the \CaIR~\AA\ observations were most often carried out in non-polarimetric mode. 

\begin{figure*}[!ht]
\includegraphics[bb = 0 17 482 161,width=\textwidth]{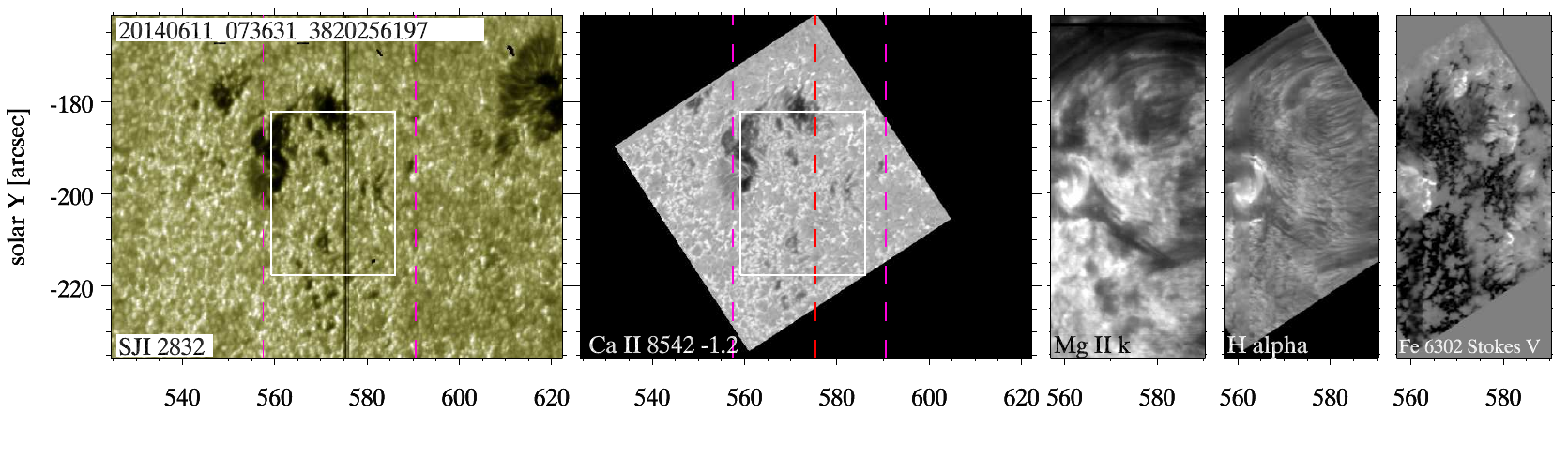}
\includegraphics[bb = 0 18 482 161, width=\textwidth]{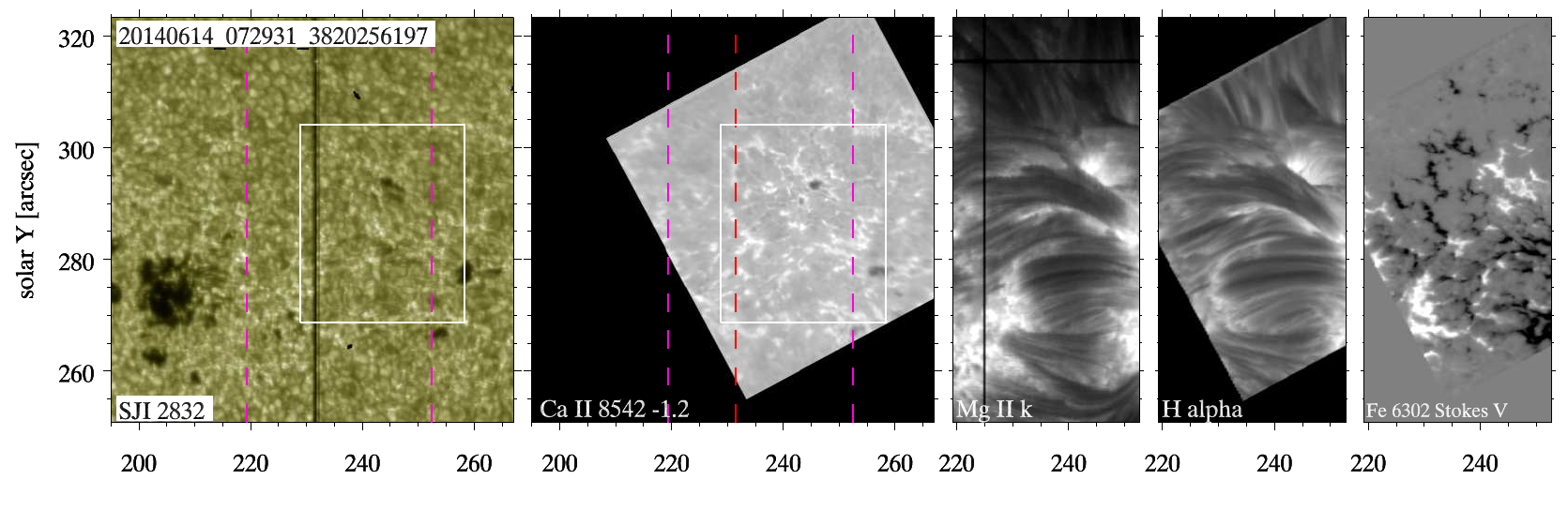}
\includegraphics[bb = 0 24 482 161, width=\textwidth]{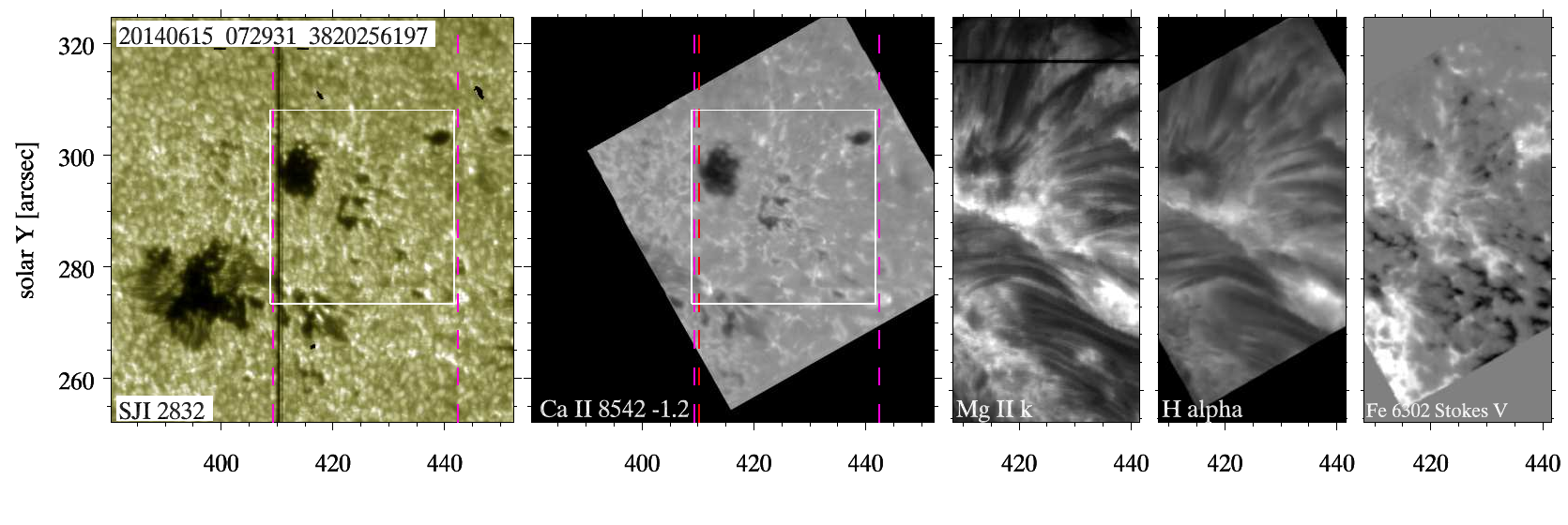}
\includegraphics[bb = 0 20 482 161, width=\textwidth]{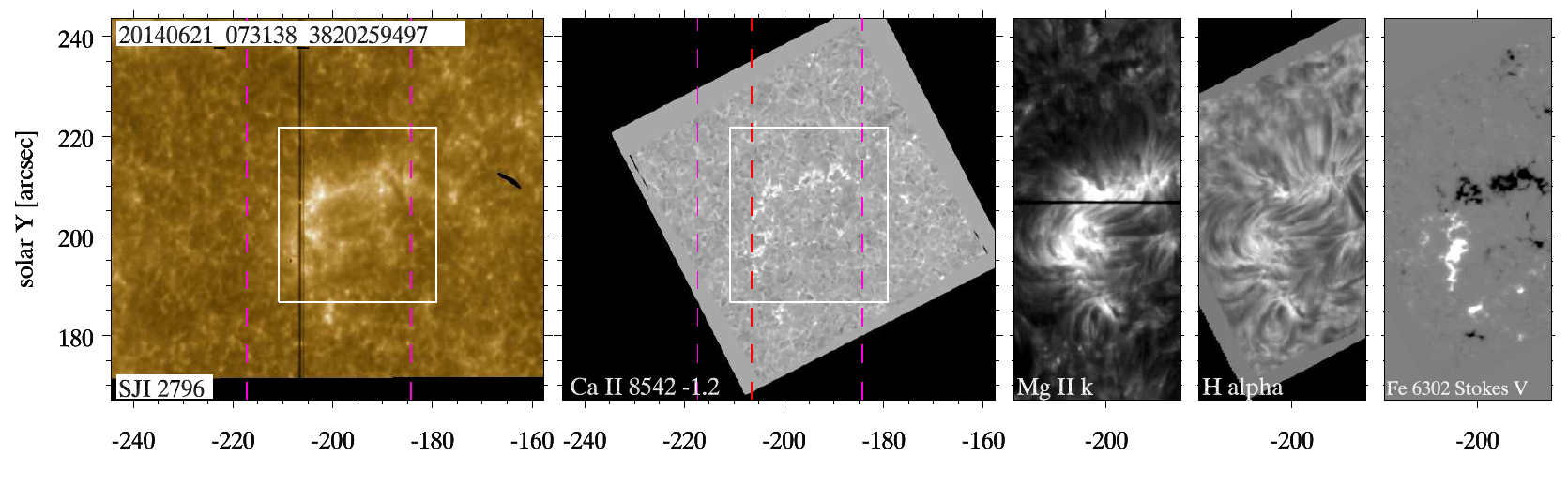}
\caption{\label{fig:raster_overview01}%
Sample images from four different datasets with large IRIS rasters. Each row shows five different diagnostics. The first two images in each row show the channel pairs that were used for IRIS and SST co-alignment and the area outlined by the white rectangle marks the region used for cross-correlation to determine offsets. The dashed red line in the second image marks the location of the IRIS slit in the SJI image to the left. The dashed purple lines mark the area covered by the IRIS raster. The three right-most images show spectroheliograms constructed from the raster data cubes: at the nominal line core wavelengths of the \ion{Mg}{ii}~k and \Halpha\ line cores, and a blue wing \FeI~6302~\AA\ Stokes V map at $-0.048$~\AA. The SST images are down-scaled to the IRIS plate scale. 
}
\end{figure*}

During the 2013 and 2014 observing seasons, photospheric spectropolarimetry was limited to one single position in the blue wing of the \FeI~6302~\AA\ line. 
The Stokes~V maps serve as effective locators of the strongest magnetic field regions and polarity indicators.
% refer to Figure
An example of such a blue wing \FeI~6302~\AA\ Stokes~V map can be seen in Fig.~\ref{fig:overview02} for the 2014-Sep-09 and 2014-Sep-15 datasets, as well as in Fig.~\ref{fig:raster_overview01}. 

During later campaigns, spectral sampling of photospheric \FeI\ lines was extended. 
These observations were subjected to a fast and robust pixel-to-pixel Milne-Eddington (ME) inversion procedure. The parallel C++ implementation\footnote{\url{https://github.com/jaimedelacruz/pyMilne}} \citep{2019A&A...631A.153D} is based upon the analytical intensity derivatives described by \citet{2007A&A...462.1137O} and an efficient Levenberg-Marquardt algorithm that is described in \citet{2019A&A...623A..74D}.

Example line of sight (LOS) magnetic field strength ($B_\textrm{LOS}$) maps from \FeI~6173~\AA\ inversions are shown in Fig.~\ref{fig:overview02} for observing date 2015-Jun-26 and in Fig.~\ref{fig:overview03} for 2015-Sep-17. 
The database contains maps of $B_\textrm{LOS}$, plane of the sky magnetic field strength $B_\textrm{perp}$, and LOS velocity from these ME inversions. 

For datasets for which spectropolarimetric \CaIR~\AA\ observations were taken, we include magnetograms that were constructed by summing Stokes~V data from the blue wing of the \ion{Ca}{ii} line, and subtracting the corresponding sum from the red wing.
These serve as photospheric magnetic field maps in a similar way as the \FeI~6302~\AA\ Stokes~V maps.
Examples can be found in Fig.~\ref{fig:overview03} for observing dates 2016-Apr-29 and 2016-Sep-03.

\subsection{IRIS and SST co-alignment}
\label{sec:coalign}
For the co-alignment of the IRIS and SST data, we employ cross-correlation of image pairs that are morphologically as similar as possible. 
Most often, the SJI 2796 and \CaIR~\AA\ wing (at 0.8 -- 1.2~\AA\ offset from line core) or \CaK\ wing show similar enough scenes to give satisfying results. 
This is particularly true for more quiet regions with the characteristic mesh-like pattern from acoustic shocks and the surrounding network of high-contrast bright regions. 
For active regions with enhanced flaring activity or large sunspots, the SJI 2796 and \CaIR~\AA\ wing pair can have more dissimilar scenes and therefore the co-alignment can be less reliable. 

The combination SJI~2832 with CRISP WB or \Halpha\ far wing gives excellent co-alignment results since both channels show pure photospheric scenes. 
However, SJI~2832 is not always selected for the IRIS observing programs to limit the data rate and improve the cadence of the other SJI channels.

\begin{figure*}[!ht]
\includegraphics[bb = 0 15 482 148, width=\textwidth]{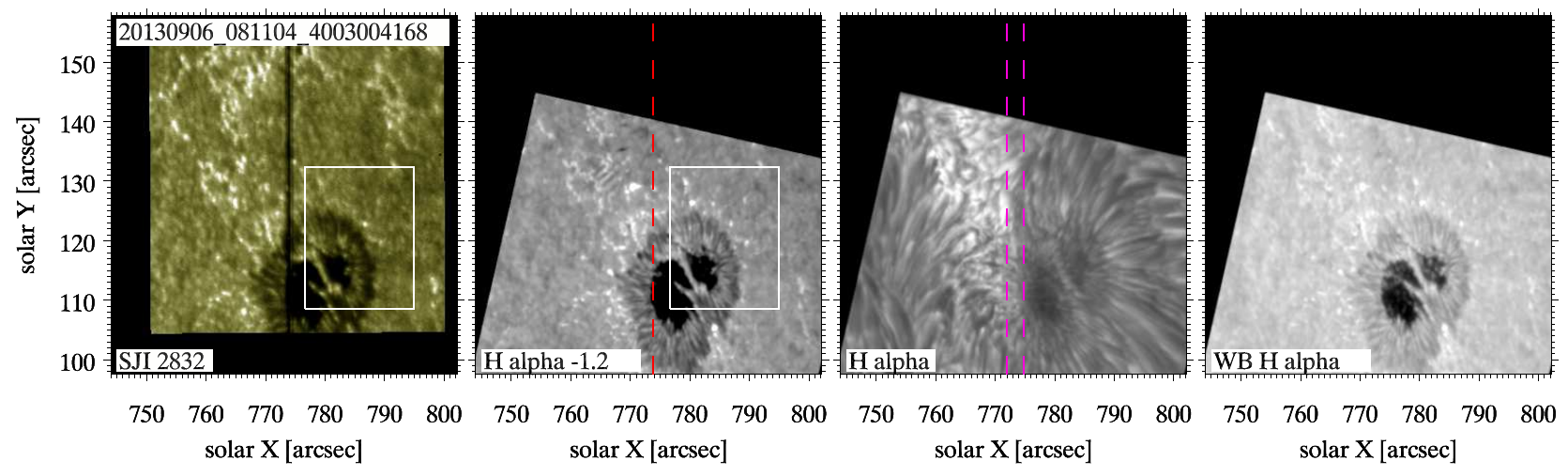}
\includegraphics[bb = 0 15 482 148, width=\textwidth]{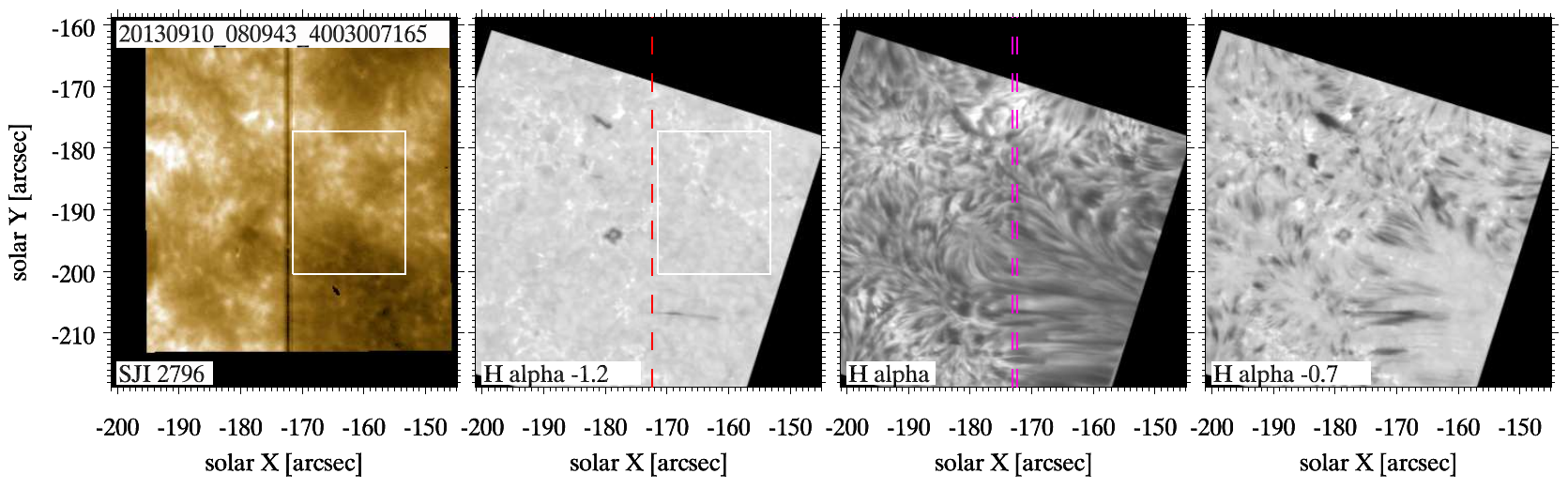}
\includegraphics[bb = 0 18 482 148, width=\textwidth]{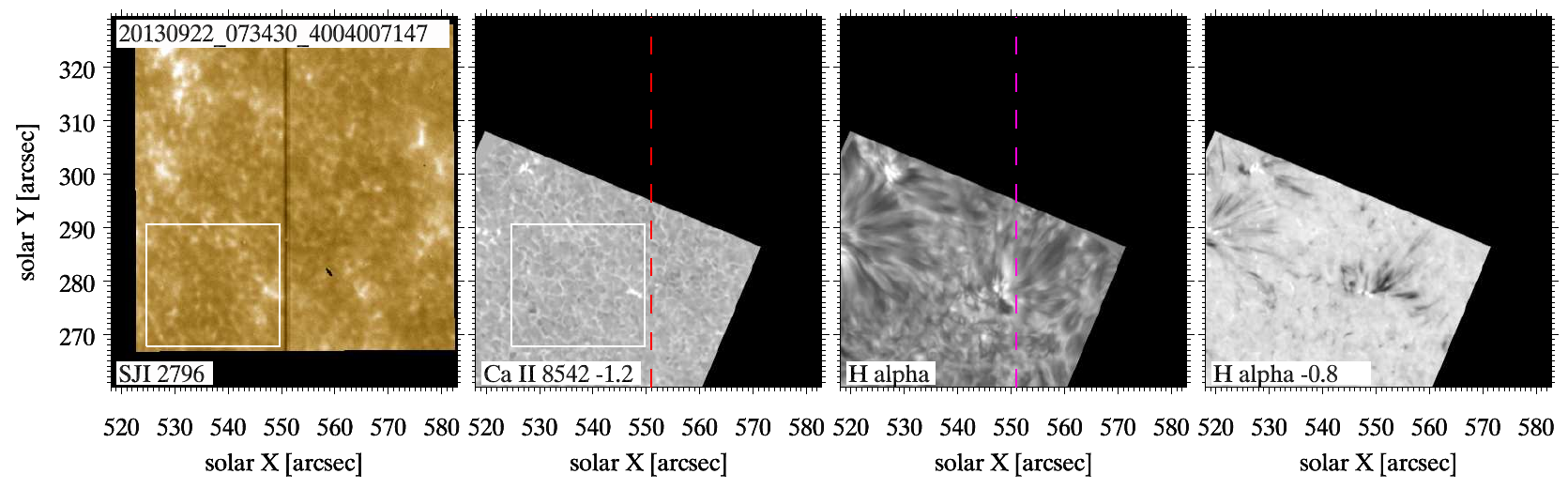}
\includegraphics[width=\textwidth]{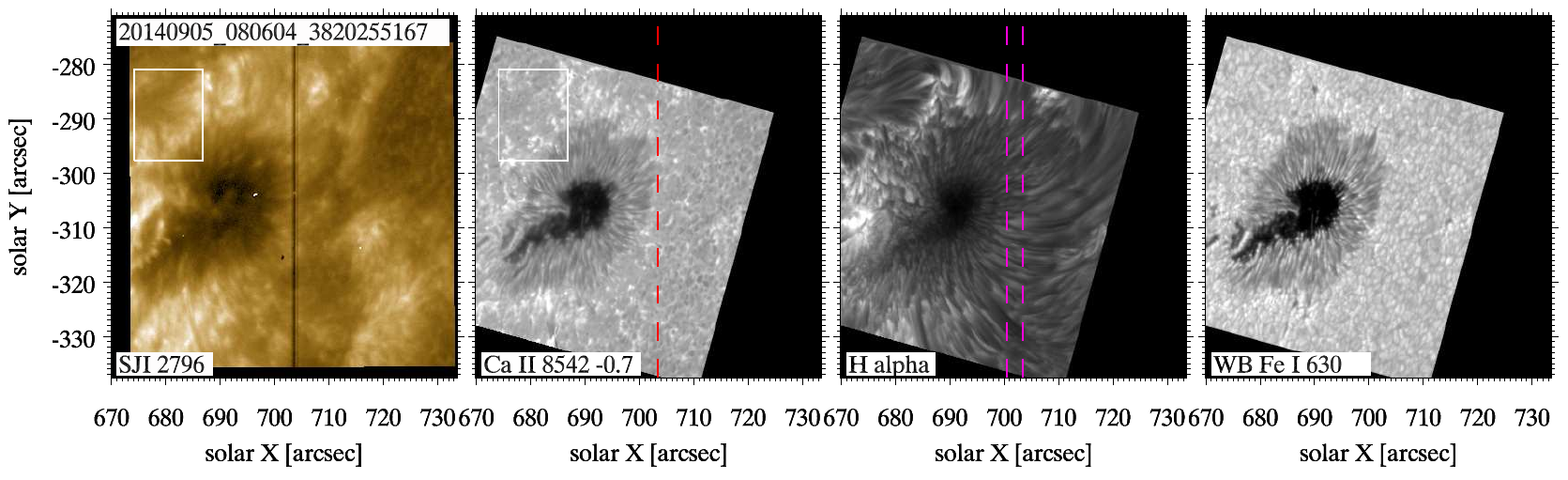}
\caption{\label{fig:overview01}%
Sample images from four different datasets. Each row shows four different diagnostics. The first two images on each row show the channel pairs that were used for IRIS and SST co-alignment and the area outlined by the white rectangle marks the region used for cross-correlation to determine offsets. The dashed red line in the second image marks the location of the IRIS slit in the SJI image to the left. The dashed purple lines in the third image mark the area covered by the IRIS raster. The SST images are down-scaled to the IRIS plate scale. 
}
\end{figure*}

\begin{figure*}[!ht]
\includegraphics[bb = 0 15 482 148, width=\textwidth]{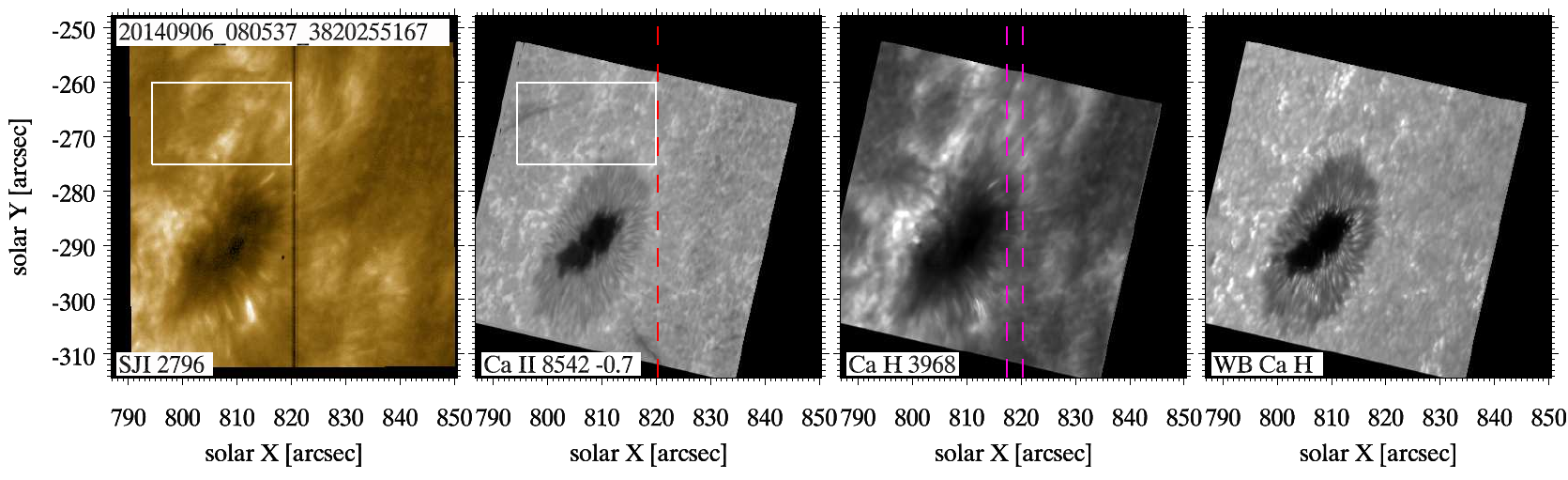}
\includegraphics[bb = 0 15 482 148, width=\textwidth]{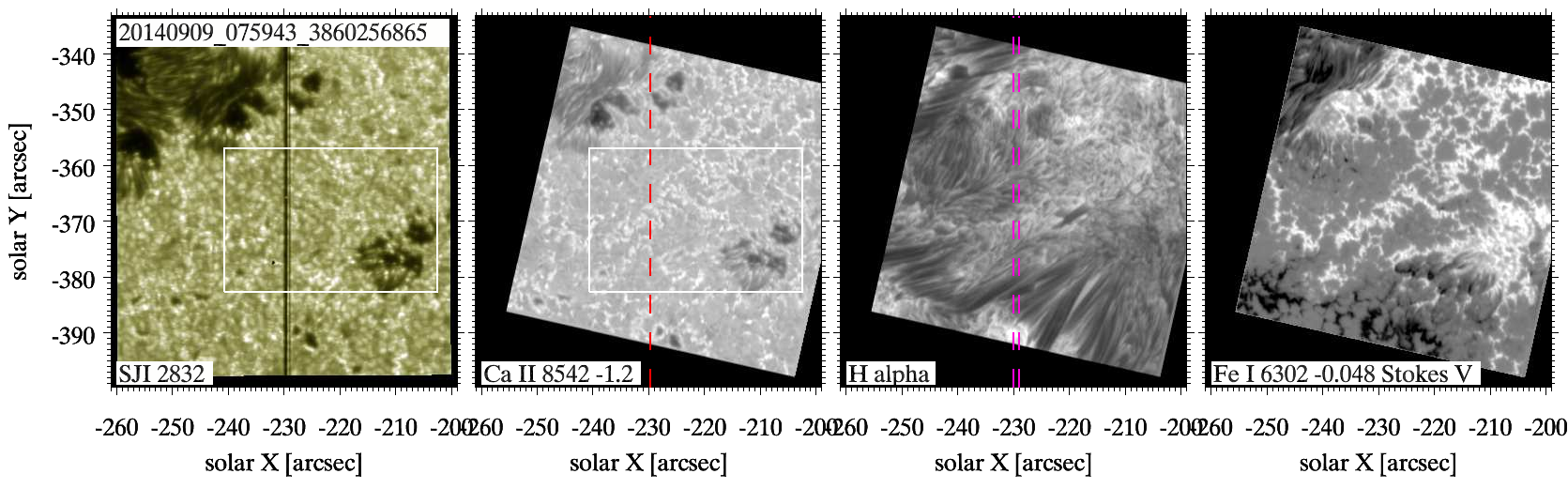}
\includegraphics[bb = 0 17 482 148, width=\textwidth]{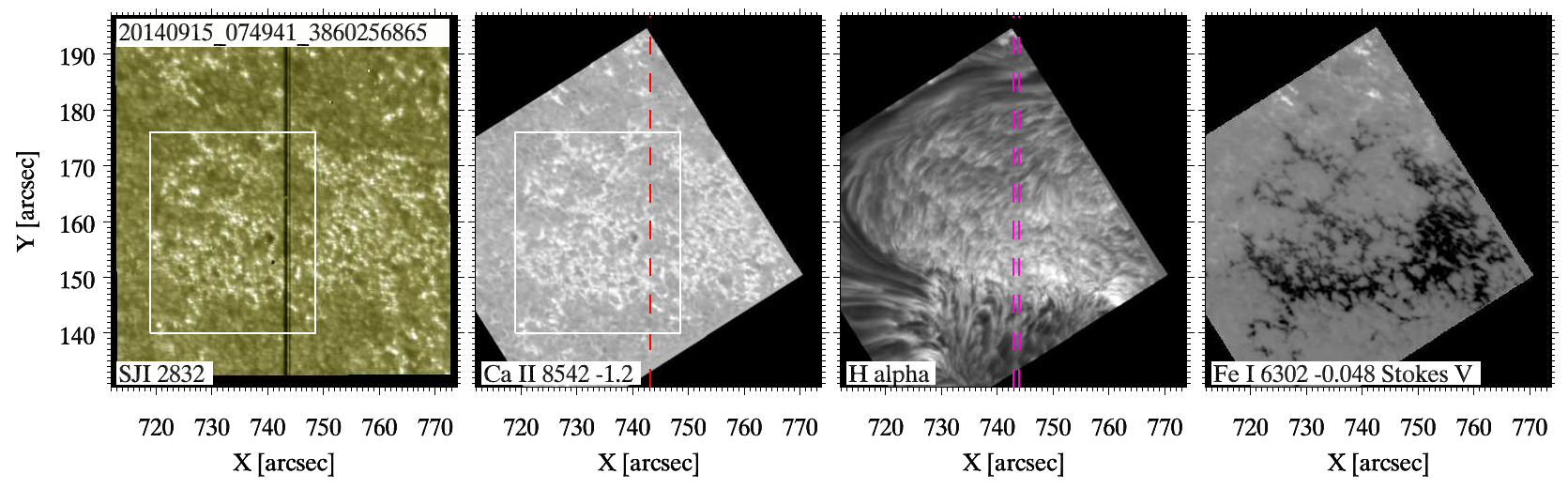}
\includegraphics[width=\textwidth]{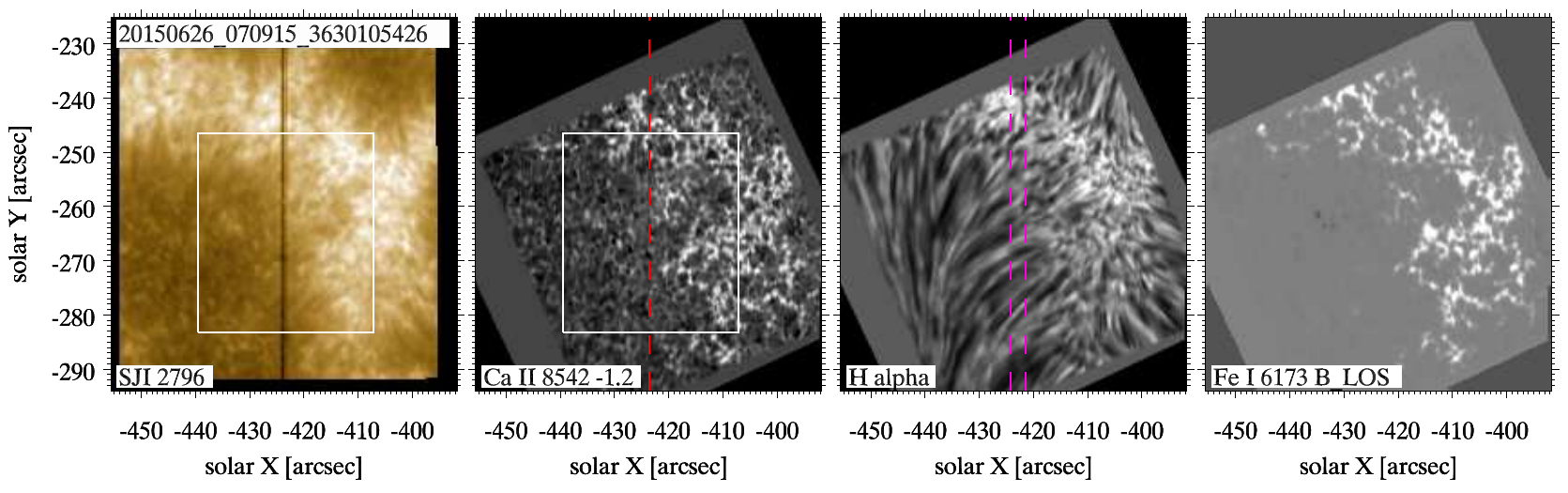}
\caption{\label{fig:overview02}%
Sample images from four different datasets. The format is the same as Fig.~\ref{fig:overview01}.
}
\end{figure*}

\begin{figure*}[!ht]
\includegraphics[bb = 0 23 482 148, width=\textwidth]{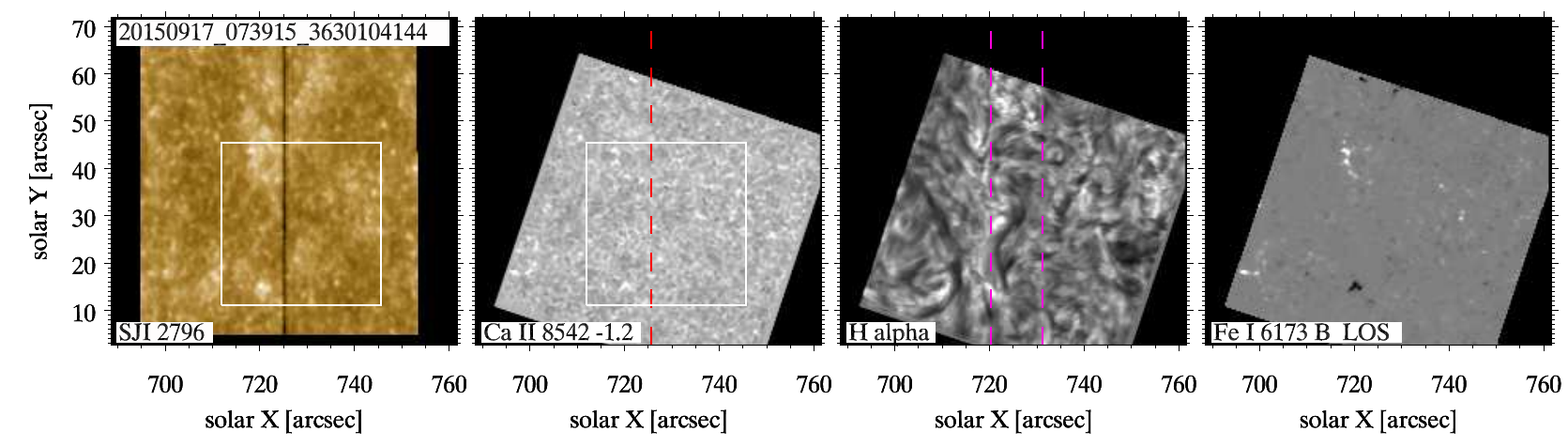}
\includegraphics[bb = 0 21 482 148, width=\textwidth]{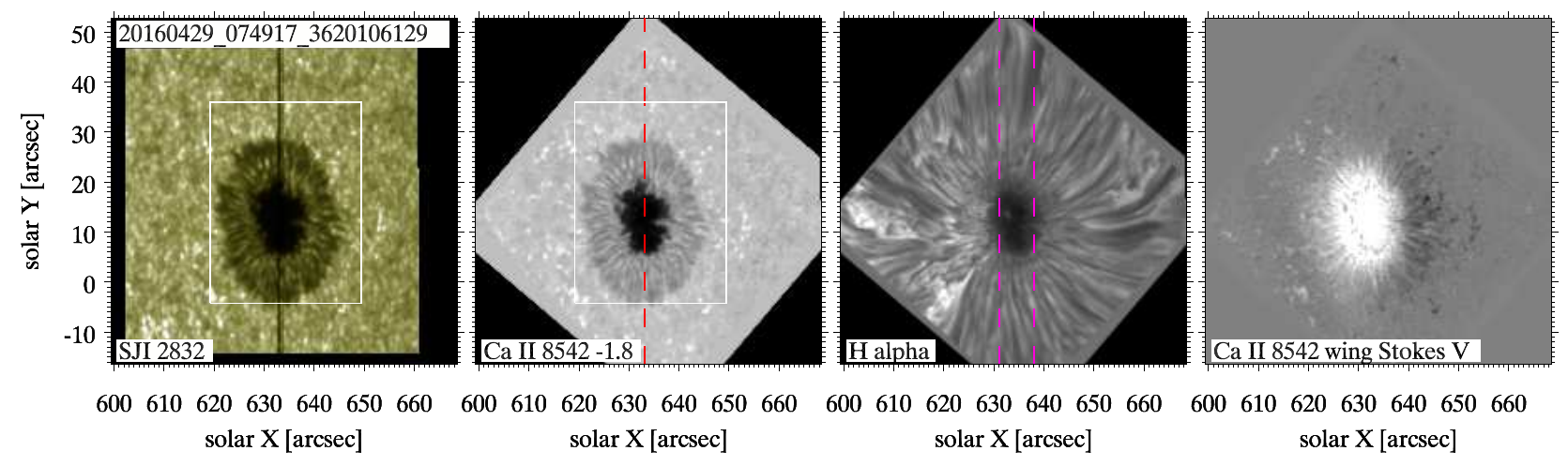}
\includegraphics[bb = 0 18 482 148, width=\textwidth]{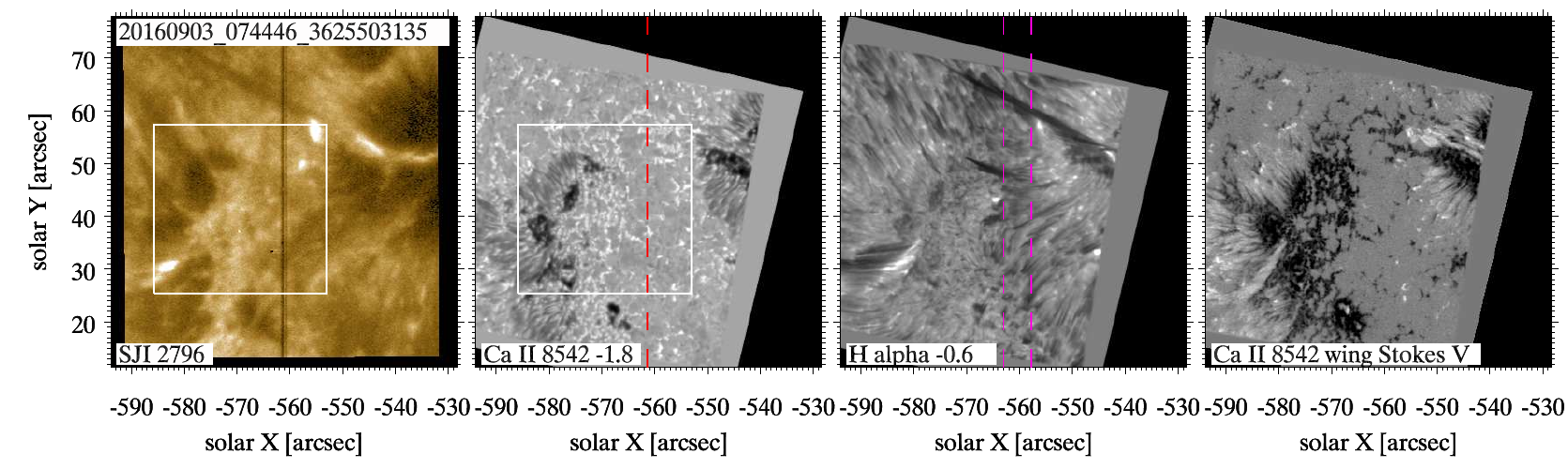}
\includegraphics[width=\textwidth]{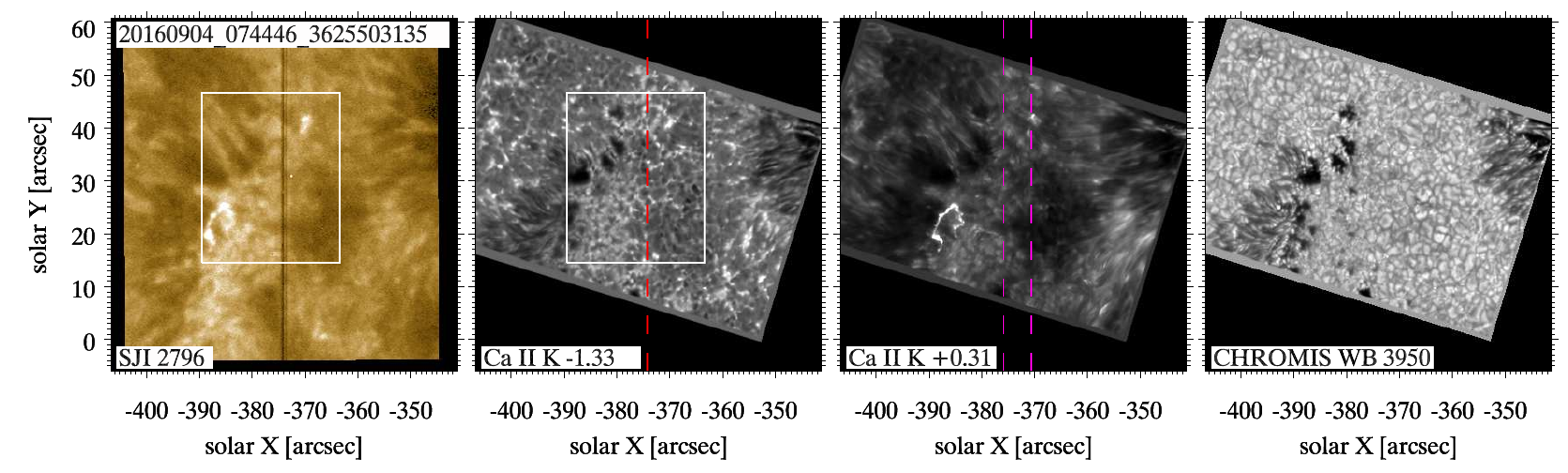}
\caption{\label{fig:overview03}%
Sample images from four different datasets. The format is the same as Fig.~\ref{fig:overview01}.
}
\end{figure*}

Before cross-correlation, the plate scales between image pairs are matched.
Offsets are then determined by cross-correlation over a subfield of the common FOV of image pairs that are closest in time. 
Examples of such subfields are outlined by white rectangles in Figs.~\ref{fig:overview01}--\ref{fig:raster_overview01}.
The raw offsets are then smoothed with a temporal window to account for jitter due to noise. 
The offsets that are applied to the data are interpolated to the relevant time grid of the particular diagnostic. 

The precision of the alignment is limited by a number of factors. 
Formation height differences between the diagnostics used for cross-correlation may introduce a systematic offset that is difficult to account for. 
This is probably of limited concern for cross-correlation between photospheric diagnostics involving SJI~2832, but it is more uncertain between SJI~2796 and the \CaIR\ wing. 
The systematic offset may be higher for oblique observing angles towards the limb and may also depend on the type of target (for example, active regions with flaring activity that appears less prominent in the \CaIR\ wing than in the \ion{Mg}{ii}~k core).
Furthermore, varying seeing conditions at the SST inevitably lead to image distortions that cannot be fully accounted for in post-processing. 
We estimate that the error in the co-alignment can be as good as or better than one IRIS pixel (0\farcs17, in the case of SST data taken under excellent conditions and closely matching diagnostic pairs in the cross-correlation). 
However, we also see local offsets due to image warping that can be as large as $\sim$2 IRIS pixels. 
These local offsets vary in magnitude proportionally with the seeing conditions. 

For the current release of data to the database, the IRIS data is kept as reference. 
This means that the SST data is down-scaled to the IRIS plate scale (for CRISP with a factor 2.9, for CHROMIS with a factor 4.4), rotated and clipped to match the IRIS FOV and orientation, and clipped in time to match the IRIS observation duration. 
We have also applied the reverse approach, keeping at least the SST data at its superior spatial resolution for analysis in earlier publications. 
These types of data products are considered for future data releases but we note that the quality control is a laborious effort, partly due to the alignment uncertainties outlined above. 

For future studies one can consider the use of the more highly resolved SST data to
uncover possible fine structure below the IRIS resolution, which could be of importance for the interpretation of the data. Such analysis could be performed by comparing individual spectra from both datasets, or by using a newly developed spatially coupled inversion method
that allows for the combining of datasets acquired at different spatial resolution \citep{2019A&A...631A.153D}. 
Each dataset would set constraints in the reconstructed model down to the smallest spatial scales that are present in the data without affecting the information provided by the other datasets that are included in the inversion.

\section{Data in the database}
\label{sec:database}

Table~\ref{table:datasets} gives an overview of various parameters that characterize datasets in the public database. 
There is a variety of targets, including the quiet Sun, coronal holes, enhanced networks, active regions with and without sunspots, and plages. 

The data can be accessed through the database, available through the public web portal at the IRIS web pages at 
LMSAL\footnote{\url{https://iris.lmsal.com/search/}}.
The data will also be available through the Hinode Science Data Centre Europe hosted at the University of Oslo\footnote{\url{http://sdc.uio.no/sdc/}}.
The data products that are publicly released are FITS files in so-called IRIS level~3 format. 
Level~3 data are data cubes that are a recast of the standard IRIS level~2 data files. 
Level~2 data are the science-ready data files that have been processed to include corrections for dark current and flat field, geometric distortions and wavelength calibration
\citep{2018SoPh..293..149W}. % Wuelser+ IRIS calibration 
The level~3 data are four-dimensional data cubes with $(x,y,\lambda,t)$-axes: the spatial $x$-axis along the raster slit positions, the spatial $y$-axis along the spectrograph slit, the $\lambda$-axis along the wavelength dimension, and $t$ the temporal axis.  

% https://www.crispex.org - short usage tutorial 
The level~3 data cubes in the database can be readily accessed with CRISPEX
\citep{2012ApJ...750...22V, %Vissers & Rouppe van der Voort CRISPEX
2018arXiv180403030L}, % Lofdahl+ CHROMISRED
a graphical user interface written in the Interactive Data Language (IDL).
It allows for side-by-side browsing and basic time-series analysis of the IRIS
rasters, SJIs and CRISP or CHROMIS data.
The CRISPEX interface is distributed as part of the IRIS package in SolarSoft IDL and can also
be downloaded\footnote{\url{ https://github.com/grviss/crispex}\label{fnote:crispex}} 
separately; however, it requires SolarSoft for full functionality when inspecting the IRIS-SST
data.
Tutorials for its use are available
online.\textsuperscript{\ref{fnote:crispex},}\footnote{\url{https://iris.lmsal.com/tutorials.html}}

%===========================================================================
\begin{sidewaystable*}
%\begin{table*}[bth]
\caption{Overview of the datasets available in the database.}
\centering
\begin{tabular}{cccccccccccccc}%
        \hline \hline
Date\tablefootmark{a} & Time\tablefootmark{b} & OBSID\tablefootmark{c} & \multicolumn{4}{c}{Raster\tablefootmark{d}} & Target\tablefootmark{e} & Pointing\tablefootmark{f} & $\mu$\tablefootmark{g} & \multicolumn{3}{c}{SST} & Ref.\tablefootmark{k} \\
 & & & $n$ & type & FOV [\arcsec] & Cad. [s] & & (solar $X,Y$) [\arcsec] & & Overlap\tablefootmark{h} & diagnostics\tablefootmark{i} & Cad.\tablefootmark{j} [s] & \\
\hline
20130906 & 081104 & 4003004168 & 4 & sparse & $3 \times 50$ & 11.5 & AR, S & 773, 128 & 0.57 & 00:45:30 & \Halpha & 5.5 & 6 \\
20130910 & 080943 & 4003007165 & 2 & sparse & $1 \times 50$ & 10.0 & AR & $-173, -189$ & 0.96 & 01:12:54 & \Halpha & 5.5 & 1, 2 \\
20130922 & 073430 & 4004007147 & 1 & s\&s & $0.3 \times 60$ & 4.2 & CH & 551, 295 & 0.77 & 02:02:16 & \Halpha, \ion{Ca}~8542, \ion{Fe}~6302 & 10.9 & 1, 2, 3 \\
20140611 & 073631 & 3820256197 & 96 & dense & $31 \times 175$ & 516 & AR & $573, -202$ & 0.77 & 02:50:38 & \Halpha, \ion{Ca}~8542, \ion{Fe}~6302 & 11.4 & 4, 8 \\
20140614 & 072931 & 3820256197 & 96 & dense & $31 \times 175$ & 516 & AR & 235, 277 & 0.92 & 00:51:32 & \Halpha, \ion{Ca}~8542, \ion{Fe}~6302 & 11.4 & 6 \\
20140615 & 072931 & 3820256197 & 96 & dense & $31 \times 175$ & 516 & AR & 425, 278 & 0.84 & 01:02:36 & \Halpha, \ion{Ca}~8542, \ion{Fe}~6302 & 11.4 & 6 \\
20140621 & 073138 & 3820259497 & 96 & dense & $31 \times 175$ & 908 & QS & $-$201, 258 & 0.94 & 01:43:41 & \Halpha, \ion{Ca}~8542, \ion{Fe}~6302 & 11.5 & 9 \\
20140905 & 080604 & 3820255167 & 4 & sparse & $3 \times 60$ & 13.2 & AR, S & 702, $-304$ & 0.59 & 01:53:00 & \Halpha, \ion{Ca}~8542, \ion{Fe}~6302 & 11.6 & 5 \\
20140906 & 080537 & 3820255167 & 4 & sparse & $3 \times 60$ & 13.2 & AR, S & 819, $-281$ & 0.41 & 02:01:02 & \Halpha, \ion{Ca}~8542, \ion{Fe}~6302 & 11.6 & 5 \\
 & & & & & & & & & & 02:00:53 & \CaH & 11.6 & 5 \\
20140909 & 075943 & 3860256865 & 4 & dense & $1 \times 60$ & 21 & AR & $-230, -366$ & 0.89 & 01:50:46 & \Halpha, \ion{Ca}~8542, \ion{Fe}~6302 & 11.6 & 8 \\
20140915 & 074941 & 3860256865 & 4 & dense & $1 \times 60$ & 21 & AR & 743, 164 & 0.60 & 01:16:21 & \Halpha, \ion{Ca}~8542, \ion{Fe}~6302 & 11.6 & 8 \\
20150626 & 070915 & 3630105426 & 8 & dense & $2.3 \times 60$ & 25 & QS & $-$422, $-$260 & 0.85 & 02:31:06 & \Halpha, \ion{Ca}~8542, \ion{Fe}~6173 & 16.1 & 10 \\
20150917 & 073915 & 3630104144 & 32 & dense & $10.2 \times 60$ & 99 & QS & 725, 38 & 0.65 & 00:24:46 & \Halpha, \ion{Ca}~8542, \ion{Fe}~6173 & 24.1 & 7 \\
20160429 & 074917 & 3620106129 & 8 & sparse & $7 \times 60$ & 41 & AR, S & 634, 18 & 0.74 & 01:30:09 & \Halpha, \ion{Ca}~8542 & 20 & 14, 16 \\
20160903 & 074446 & 3625503135 & 16 & dense & $5 \times 60$ & 21 & AR & $-561, 44$ & 0.81 & 02:14:30 & \Halpha, \ion{Ca}~8542 & 20 & 11, 15 \\
 & & & & & & & & & & 00:34:20 & \CaK & 13 & 12, 13 \\
20160904 & 074446 & 3625503135 & 16 & dense & $5 \times 60$ & 21 & AR & $-374, 27$ & 0.91 & 01:09:36 & \Halpha, \ion{Ca}~8542 & 20 & 15 \\
 & & & & & & & & & & 00:34:04 & \CaK & 22 & 12 \\
\end{tabular}
\tablefoot{%
\tablefoottext{a}{Observing date in format year, month, day.} \\
\tablefoottext{b}{Starting time (UT) of observations in format hour, min, s.} \\
\tablefoottext{c}{The OBSID number encodes the IRIS observing configuration in a unique number \citep[see Tables~12--14 in][]{2014SoPh..289.2733D}. % BDP IRIS
The combination <\textit{Date}>\_<\textit{Time}>\_<\textit{OBSID}> constitutes a unique identifier to the dataset. 
} \\
\tablefoottext{d}{The IRIS spectrograph slit covers a region on the Sun through a raster of $n$ slit positions, with a separation of type: dense (0\farcs35),
sparse (1\arcsec), or coarse (2\arcsec). Raster type s\&s is the ``sit-and-stare'' mode for which the slit remains fixed at one location. The area covered is shown in the FOV column, and the temporal cadence in the Cad. column.} \\
\tablefoottext{e}{Target: AR: active region, QS: quiet Sun, CH: coronal hole, S: sunspot.} \\
\tablefoottext{f}{Pointing coordinates at the beginning of the time series, the target is followed by tracking solar rotation.} \\
\tablefoottext{g}{$\mu = \cos \theta$ with $\theta$ the observing angle.} \\
\tablefoottext{h}{Duration of overlap of SST observations with IRIS in format hh:mm:ss.} \\
\tablefoottext{i}{Spectral lines observed with SST. CRISP is operated independently from the instruments on the blue beam; with fixed interference filters (\CaH) or CHROMIS (\CaK). The instruments have their own cadences and overlap times and are separated in rows in the table.} \\
\tablefoottext{j}{Cadence of the SST observations.} \\
\tablefoottext{k}{References to publications based on these data sets: 
1: \citet{2014Sci...346D.315D}, % BDP+ Science twist
2: \citet{2015ApJ...799L...3R}, % Rouppe+ IRIS RBEs 
3: \citet{2015ApJ...803...44M}, % Juan+ IRIS QS grains
4: \citet{2015ApJ...809L..30C} % Mats Mg k plage
5: \citet{2015ApJ...811L..33V}, % Vissers+ PMJs
6: \citet{2015ApJ...812...11V}, % Vissers+ EBs + UVBursts
7: \citet{2016A&A...592A.100R} % Rouppe+ QSEBs
8: \citet{2016ApJ...817..124S} % Skogsrud+AR bright grains in SJI1400
9: \citet{2017A&A...597A.138R}, % Rutten contrails
10: \citet{2017Sci...356.1269M}, % Juan+ spicules Science
11: \citet{2017ApJ...850..153N}, % Daniel+ surge
12: \citet{2017ApJ...851L...6R}, % Rouppe+ plasmoids
13: \citet{2019A&A...627A.101V}, % Gregal+ CRISP+CHROMIS+IRIS EB/UVB inversions
14: \citet{2019A&A...627A..46B}, % Souvik+ sunspot atmosphere models
15: \citet{2020A&A...633A..58O} % Ada+ EBs + UVBs
16: \citet{2020A&A...638A..63D}. % Ainar & Rouppe PMJ
} \\
}
\label{table:datasets}
%\end{table*}%
\end{sidewaystable*}
% turn into sideways table? Include exposure time and raster cadence?
% A&A style guide: 
%  To indicate the omission of an entry, ellipsis dots (...) are used.
%===========================================================================

%% fig:overview 
%===========================================================================

%%===========================================================================

%% fig: raster overview
%===========================================================================
%%%%%%%%%%%%%%%%%%%%%%%%%%%%%%%%%%%%%%%%%%%%%%%%%%%%
\begin{acknowledgements}
This paper is dedicated to Ted Tarbell who passed away in April 2019. Ted was the leader of the LMSAL SVST/SST campaigns since the 1980s through the 2000s and participated with great enthusiasm in the LMSAL campaigns in 2013 and 2014. Ted was a great friend and an inspiring mentor to his junior colleagues.
The Swedish 1-m Solar Telescope is operated on the island of La Palma
by the Institute for Solar Physics of Stockholm University in the
Spanish Observatorio del Roque de los Muchachos of the Instituto de
Astrof{\'\i}sica de Canarias.
The Institute for Solar Physics is supported by a grant for research infrastructures of national importance from the Swedish Research Council (registration number 2017-00625).
IRIS is a NASA small explorer mission developed and operated by LMSAL with mission operations executed at NASA Ames Research Center and major contributions to downlink communications funded by ESA and the Norwegian Space Centre.
We thank the following people for their assistance 
at the SST:
Jack Carlyle, 
Tiffany Chamandy, 
Henrik Eklund, 
Thomas Golding, 
Chris Hoffmann,
Charalambos Kanella, 
Ingrid Marie Kjelseth,  
and Bhavna Rathore.
We further acknowledge excellent support at the SST by Pit S{\"u}tterlin. 
% IRIS planners
We are also grateful to 
%Paul Boerner, Sean McKillop, Charles Kankelborg, Joten Okamoto, Ted Tarbell and Sarah Jaeggli as 
the IRIS planners for the IRIS-SST coordination.
% SOLARNET > Hinode Science Datacenter UiO
This project has received funding from the European Union's Horizon 2020 research and innovation programme under grant agreement No 824135.
% RoCS and Toppforsk:
This research is supported by the Research Council of Norway, project number 250810, and through its Centres of Excellence scheme, project number 262622.
% Bart and LMSAL
BDP and colleagues at LMSAL and BAERi acknowledge support from NASA contract NNG09FA40C (IRIS).
% Jaime: 
JdlCR is supported by grants from the Swedish Research Council (2015-03994), the Swedish National Space Agency (128/15) and the Swedish Civil Contingencies Agency (MSB). This project has received funding from the European Research Council (ERC) under the European Union's Horizon 2020 research and innovation programme (SUNMAG, grant agreement 759548).
% Vasco
VMJH and SJ receive funding from the European Research Council (ERC) under the European Union’s Horizon 2020 research and innovation programme (grant agreement No. 682462).
% Juan:
JMS is supported by NASA grants NNX17AD33G, 80NSSC18K1285, 
%and NNG09FA40C (IRIS), % see above
and NSF grant AST1714955. 
% Gregal:
GV is supported by a grant from the Swedish Civil Contingencies Agency (MSB).
% Clara:
CF acknowledges funding from CNES.
We made much use of NASA's Astrophysics Data System Bibliographic Services.
\end{acknowledgements}

%\bibliographystyle{aa-note}
%\bibliography{qseb} 
%\bibliography{iris_sst} 

%\begin{appendix}
%\section{}
%\label{app:}
%\end{appendix}

\end{document}